\documentclass[10pt,twocolumn,floatfix,reprint]{revtex4-2}

\usepackage{graphicx,amssymb,amsmath,hyperref}
\usepackage{xfrac}
\usepackage{courier}
\usepackage{booktabs}
\usepackage{ragged2e} 

\begin{document}

\title{Fragility of Chess positions:\\ measure, universality and tipping points}
%


\author{Marc Barthelemy$^{1,2}$}
\affiliation{$^1$ Universit\'e Paris-Saclay, CNRS, CEA, Institut de Physique Th\'eorique, 91191, Gif-sur-Yvette, France}
\affiliation{$^2$ Centre d’Analyse et de Math\'ematique Sociales (CNRS/EHESS) 54 Avenue de Raspail, 75006 Paris, France}

\begin{abstract}

  We introduce a novel metric to quantify the fragility of chess positions using the interaction graph of pieces. This fragility score \(F\) captures the tension within a position and serves as a strong indicator of tipping points in a game. In well-known games, maximum fragility often aligns with decisive moments marked by brilliant moves. Analyzing a large dataset of games, we find that fragility typically peaks around move $15$, with pawns ($\approx 60\%$) and knights  ($\approx 20\%$) frequently involved in high-tension positions. Comparing the Stockfish evaluation with the fragility score, we observe that the maximum fragility ply often serves as a critical turning point, where the moves made afterward can determine the outcome of the game. Remarkably, average fragility curves show a universal pattern across a wide range of players, games, and openings, with  a subtle deviation observed in games played by the engine Stockfish. Our analysis reveals a gradual buildup of fragility starting around $8$ moves before the peak, followed by a prolonged fragile state lasting up to $15$ moves. This suggests a gradual intensification of positional tension leading to decisive moments in the game. These insights offer a valuable tool for both players and engines to assess critical moments in chess.
  
\end{abstract}

\maketitle




\section{Towards a quantitative theory of the chess game}

Richard R\'eti, one of the world's top players in the early 20th century and a leading proponent of hypermodernism in chess (alongside Nimzovich), gave a series of lectures on chess in Buenos Aires in 1924. These lectures were published in Spanish \cite{Reti} under a title that could be translated as `Scientific Lectures on Chess'. In the preface, R\'eti outlines three stages toward developing a scientific understanding of chess. The first stage involves the empirical collection of data, the second focuses on constructing typologies, and the third aims to establish laws based on these typologies. While this ambitious program remains unfulfilled, a parallel approach, driven largely by advances in computer science, has since come to dominate the field. Prominent scientists such as Turing \cite{Turing}, Shannon \cite{Shannon:1950}, Simon, Newell \cite{Newell} and others discussed about 70 years ago the importance of chess as a prototype for testing algorithms and artificial intelligence (see the compendium \cite{Levy:1988}). Indeed, the study of chess offers a rich intersection between computational science and complex systems analysis. With its simple rules yet vast strategic depth, chess provides an ideal platform for developing and testing algorithms in artificial intelligence (AI), machine learning, and decision theory. Its deterministic nature allows for the application of various computational techniques, from heuristic searches to neural networks, which offer insights into optimization and problem-solving. Chess has historically been central to AI research, exemplified by landmark events like IBM’s Deep Blue defeating Garry Kasparov in 1997, a milestone in AI’s capabilities \cite{Sadler:2019}. More recent innovations, such as AlphaZero, further highlight the game's importance in advancing computational methods \cite{Silver:2018}. Chess also extends beyond algorithms, introducing psychological complexity as players engage in predicting and concealing strategies, adding layers of depth to the game.

However, chess, when viewed as a relatively simple complex system governed by the interactions of a small number of constituents, could serve as a testbed for many ideas developed in complex systems science. Mathematically, chess can be represented as a decision tree where each branch leads to a win, loss, or draw, and the challenge lies in selecting the best move amid the vast combinatorial complexity, especially during the middlegame, where the goal is to navigate toward favorable branches. It is thus surprising that complexity science—and in particular, statistical physics—has had relatively little to say about this system. However, the recent rise of online chess platforms has enabled large-scale data analysis, changing this situation to some extent and allowing for the gradual introduction of concepts and tools from statistical physics and complex systems \cite{Perotti:2013, DeMarzo:2022, Blasius:2009, Maslov:2009, Ribeiro, Schai:2014, Schai:2016, Atash, Chow:2023, Barthelemy2023}. Researchers have identified patterns such as power-law distributions in opening move frequencies, reflecting the self-similar nature of the game tree \cite{Blasius:2009, Maslov:2009}. Long-range memory effects in game sequences, which vary according to player skill levels, have also been observed \cite{Schai:2014, Schai:2016} and the impact of chess experts' knowledge was discussed in \cite{Chassy:2011}. Additionally, the response time distribution in rapid chess was analyzed in \cite{Sigman}, and the decision-making process of chess players was examined in \cite{Chacoma}.

Chess is traditionally divided into three phases: the opening, middlegame, and endgame, with extensive study dedicated to each. A key focus in chess discussions is on `critical points' (also referred to as 'turning' or 'tipping points')—positions where the choice of move can significantly influence the outcome of the game (see, for example, \cite{Dorfman}). These critical positions are often unstable, and a small mistake can lead to dramatic shifts in the game's trajectory. One of the enduring challenges in chess is identifying the best move, particularly in the middlegame when the combinatorial possibilities are vast. In this study, we address the problem of identifying these pivotal positions that largely determine the course of the game. The existence of such turning (or tipping) points is connected to how rapidly a position can transform, often in just a few moves: material (im)balance, pawn structure, and square control can all undergo significant changes within a short span. This concept relates to the `fragility' of a position, which we aim to quantify by analyzing piece interactions through graph theory. We define fragility based on the betweenness centrality of pieces under attack. Our objective is to calculate a fragility score for each position, aggregate these scores across multiple games, and analyze the statistical properties of fragility over time. The paper is structured as follows: we first define the interaction graph of a position, which allows us to quantify its fragility and to define a fragility score. We then apply this metric to different games, demonstrating that it can identify turning points and, interestingly, reveals a universal pattern when averaged over a large number of games.

\section{Characterizing the fragility of a position}

\subsection{The interaction graph}

\begin{figure}[h!]
    \centering 
    \begin{minipage}{0.3\textwidth} 
        \centering 
        \includegraphics[width=\textwidth]{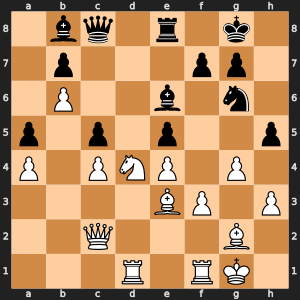}
        \begin{picture}(0,0)
            \put(-110,150){\textbf{(a)}}
        \end{picture}
    \end{minipage}
    
    \vspace{0.2cm} 

    \begin{minipage}{0.45\textwidth} 
        \centering 
        \includegraphics[width=\textwidth]{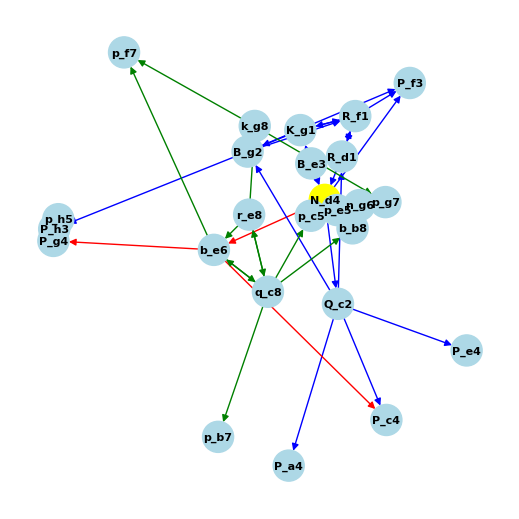}
        \begin{picture}(0,0)
            \put(-110,180){\textbf{(b)}}
        \end{picture}
    \end{minipage}
    
    \justifying 
    \caption{{\bf Interaction graph.} Example on a position taken from Mehedlishvili - Van Forrest (see Data for more details). (a) Board Interaction graph at the ply $49$ after the move \texttt{Nxd4} (which happened after the excellent move \texttt{Rd4}!!). (b) The key piece is here the white knight and has the largest betweenness centrality. In this graph, we denote the corresponding pieces of nodes with uppercase letters for white pieces and lowercase for black pieces (and the position of each piece is shown on each node).}
    \label{fig:1}
\end{figure}
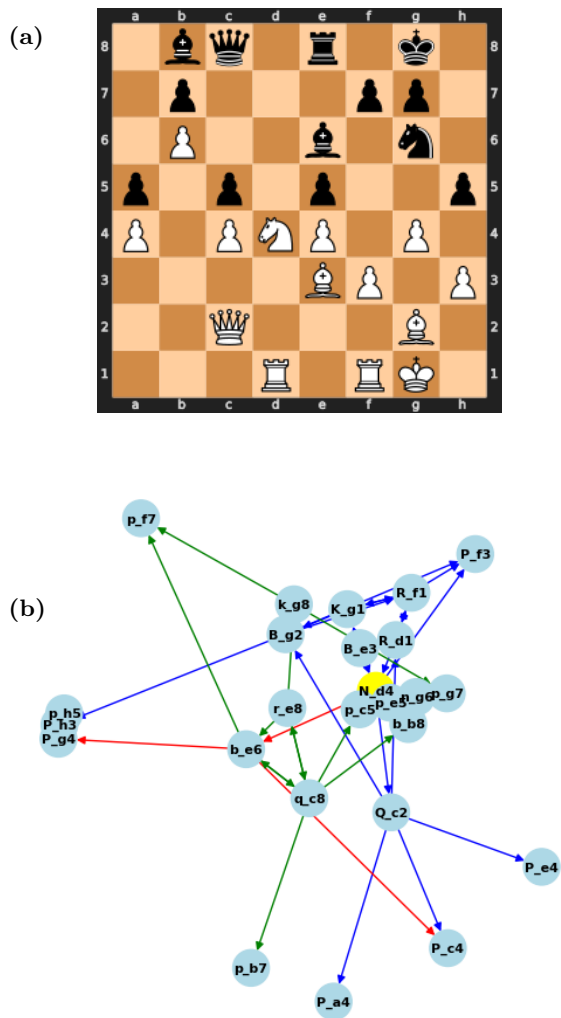

We construct the interaction graph that describes, for a given position, how the different pieces on the board attack or defend each other. More precisely, the interaction graph $G(V, E)$ consists of nodes $V$, representing the pieces on the chessboard, and directed edges $E$, representing the interactions between these pieces. Let $P = \{p_1, p_2, ..., p_n\}$ be the set of pieces on the board (both black and white), and will be the nodes of $G$ (in this graph, nodes are labelled by their pieces which are represented in uppercase for white and lowercase for black pieces). An edge $ e_{ij} \in E$ is
drawn from node $p_i$ to $p_j$ if:
\begin{itemize}
\item{} Attack interaction: $ p_i$ can legally capture $ p_j $ (in red color in Fig.~\ref{fig:1}). This edge is directed as the attack is in general not bi-directional.
\item{} Defense interaction: $ p_i $ defends $ p_j $ where both pieces are of the same color).  A defense link is then a directed edge between two pieces of the same color, indicating that one piece can move to defend another, implying it could potentially retake the opponent if the defended piece is captured (in Fig.~\ref{fig:1}, these edges are colored blue for white pieces and green for black pieces).
\end{itemize}
This graph is directed as an attack by piece $ p_i$ on piece $ p_j $ is not symmetric; $ p_i$ can attack $ p_j $, but $ p_j $ may not necessarily attack $ p_i $.  Similarly, defense interactions (where one piece defends another) are also directional, where one piece is protecting another, but the reverse is not true.
We show in Fig.~\ref{fig:1} an example of an interaction graph and the corresponding position.

An important piece in the position - namely one whose disappearance will modifiy in depth the position - is then characterized by a large number of interactions with other pieces. Centrality measures appear therefore to be very relevant for identifying key pieces in this position. In this study, we use betweenness centrality (BC) to quantify importance, as it captures how frequently a piece lies on the shortest paths between other pieces in the interaction graph. The BC $g$ of a piece $p_i$ is defined as \cite{Freeman:1977}
\begin{align}
g(p_i) = \frac{1}{(n-1)(n-2)}\sum_{s \neq p_i \neq t} \frac{\sigma_{st}(p_i)}{\sigma_{st}}
\end{align}
where $\sigma_{st} $ is the total number of shortest paths from node $s$ to node $t$, and $\sigma_{st}(p_i) $ is the number of those paths that pass through node $ p_i$, and $n$ is the total number of pieces still on the board. We used here the normalized BC in order to eliminate the effect caused by the decreasing number of pieces during the game. Additionally, since betweenness centrality measures how often a node (or piece) lies on the shortest paths between pairs of other nodes, it is essential to account for directed interactions when calculating centrality. In chess, the directed nature of these relationships is crucial, as it captures the true dynamics of the game: a piece can defend another, but this interaction is not symmetric. Thus, we compute the betweenness centrality for the directed interaction graph $G$, where edges represent directed interactions such as attacks or defenses. Calculating BC for an undirected graph would ignore this directionality, potentially losing important information about the nature of these interactions. We did test the undirected case, and while the results were similar—producing comparable values for the ply at the maximum—they were generally less precise. This outcome makes sense because, in chess, a piece A can defend a piece B, but the reverse is not always true. Since BC aims to capture cascades of exchanges, respecting the directional order is imperative.

We note that other centrality measures could be considered. Specifically, we tested the simplest one—node degree, which corresponds to the number of pieces interacting with the piece at a given node—but found that it produced less precise results and failed to highlight some key positions. We also tested eigenvector centrality, but it too yielded less precise outcomes. The superiority of betweenness centrality (BC) likely arises from the fact that it is not a local measure, accounting for more than just a single move. The BC captures cascades of exchanges, whereas degree centrality is primarily tied to a single half-move, limiting its effectiveness in identifying critical positions.

\subsection{The fragility score}

We can now define the fragility score $F$ of a position. Each piece  $p_i$ has a betweenness centrality (BC) value computed from the interaction graph corresponding to the given position. A piece with a high BC value is pivotal, as its capture could initiate a cascade of exchanges and deeply alter the structure of the position. If such a piece is under attack, the position could undergo significant changes in the next moves, making it `fragile'. Therefore, it is natural to define the fragility score of the position as
\begin{align}
  F=\sum_{p\in P} g(p)a(p)
\end{align}
where $a(p)=1$ if the piece $p$ is under attack and $a(p)=0$ otherwise. This score thus computes the total betweenness centrality of attacked pieces. If the large betweenness centrality pieces are attacked, the position is unstable and can be very dynamically changing. Here, we compute the total fragility score but note that it could be defined for white and black separately. This total fragility score represents the overall vulnerability of the position, taking into account the importance of the pieces under attack and their centrality within the interaction graph. By tracking the total fragility score throughout the game, we can identify critical moments when positions become particularly fragile, reflecting the dynamics of attack and defense.

\begin{figure}[h!]
	\centering
	\includegraphics[width=0.5\textwidth]{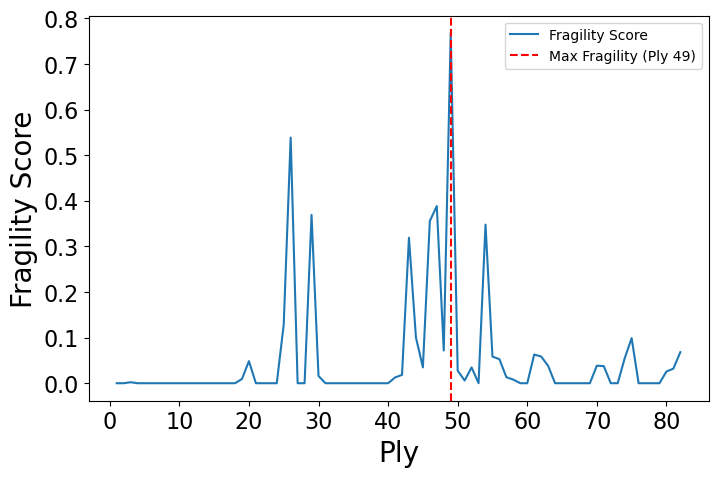}
	\caption{{\bf Fragility score}. Fragility score $F$ computed from the betweenness centrality of attacked pieces. We show here the score versus the ply for the game Mehedlishvili - Van Forrest (see Data for more details). We indicate the maximum value obtained at ply $49$ (after the move \texttt{Nxd4}) and the corresponding position is shown in Fig.~\ref{fig:1}.}
	\label{fig:2}
\end{figure}

For a single chess game, the total fragility score is computed for every half-move (or ply). Additionally, the maximum fragility
point is identified (i.e., the ply where the fragility score reaches its peak). We show an example in Fig.~\ref{fig:2} (the corresponding chessboard and interaction graph at the maximum fragility point, highlighting the position and the key pieces contributing to the fragility, are shown in Fig.~\ref{fig:1}).

Interestingly, the maximum fragility score often highlights the most critical and notable moves in a game. While we cannot conduct a large-scale comparison between human analysis and fragility predictions, we can examine specific cases. We analyzed the top 10 games ever played, as ranked by chess.com \cite{top10}, and computed the fragility score for each. For each game, we identified the move corresponding to the maximum fragility. The results are presented in the Table \ref{table1}).
\begin{table}[ht]
\centering
\begin{tabular}{|c|c|c|}
\hline
  \textbf{Game} & \textbf{Move (ply)} & \textbf{Notable} \\
   &  \textbf{at max F} &  \textbf{moves (ply)} \\ 
\hline
  Jinshi vs. Liren & \texttt{d4c3} (30) &  $40$ \\
  Chinese League 2017 (0-1) &  & \texttt{Rd4}   \\ 
  \hline
  Kasparov vs. Topalov & \texttt{Nd5} (44)   & \texttt{Rxd4} (47)\\
  Wijk aan Zee 1999 (1-0) &  &   \texttt{Re7+} (49) \\
   &  &   \texttt{Bf1} (71)\\ 
  \hline
 Morphy vs. Allies & \texttt{Ne6} (39)  &  \texttt{Nxb5} (19) \\
  Paris Opera 1858 (1-0)  &  &  \texttt{Rxd7} (25) \\
   &       &  \texttt{Qb8+} (31) \\  
  \hline
  Aronian vs. Anand & \texttt{Bg4} (33) & \texttt{Nde5} (32) \\
  Wijk aan Zee 2013 (0-1) &  &  \\
  \hline
Karpov vs. Kasparov (16) & \texttt{g5} (42)  & \texttt{g5} (42)\\
World Championship 1985 (0-1) &  &\\
  \hline
  Byrne vs. Fischer & \texttt{Nc3} (24) & \texttt{Na4} (22)\\ 
  New York 1956 (0-1) &    & \texttt{Be6} (34)\\ 
  \hline
  Short vs. Timman & \texttt{Qb4} (32)  & \texttt{Kh2} (60)\\
  Tilburg 1991 (1-0) &  & \texttt{Kg3} (63)\\
  \hline
Rotlewi vs. Rubinstein  & \texttt{Ne5} (30) & \texttt{Rxc3} (44)\\
 Lodz 1907  (0-1) &  & \texttt{Rxd2} (46)\\
\hline
  Geller vs. Euwe & \texttt{Rc1} (47)  & \texttt{Rh8} (44)\\
  Zurich 1953 (0-1) &   & \\
\hline
\end{tabular}
\caption{Selection of the ten best game of chess (according to chess.com \cite{top10}). The first column indicates the game, the second column the ply and the corresponding move leading to the maximum of the fragility score (of the game). The last column indicates the brilliant move(s) discussed in \cite{top10}.}
\label{table1}
\end{table}

In many cases, the maximum fragility coincides with brilliant and decisive moves (with minor variations of the order of a few moves), suggesting that during these high-tension phases, creativity and skill are crucial in shaping the outcome of the game. As illustrated below (Fig.~\ref{fig:4}), this phase of maximum fragility often dictates the fate of the game and naturally corresponds to critical and noteworthy moves.

\section{Statistical results for many games}

The result shown in Fig.~\ref{fig:2} is obtained for one specific game and to gain statistical insights, we extend the analysis across multiple games.
We consider here a total of $20,685$ games played by world top players: Alekhine ($1661$ games), Capablanca ($597$ games), Carlsen ($1730$ games), Fisher ($1053$), Karpov ($3079$ games), Kasparov ($4049$ games), Morphy ($211$ games), Nakamura ($686$ games), Polgar ($2398$ games), Reti ($646$ games), Spassky ($4176$ games), and also for the Stockfish engine against computers ($413$ games). The datasets are freely available from \cite{pgnmentor}.

\subsection{Maximum fragility}
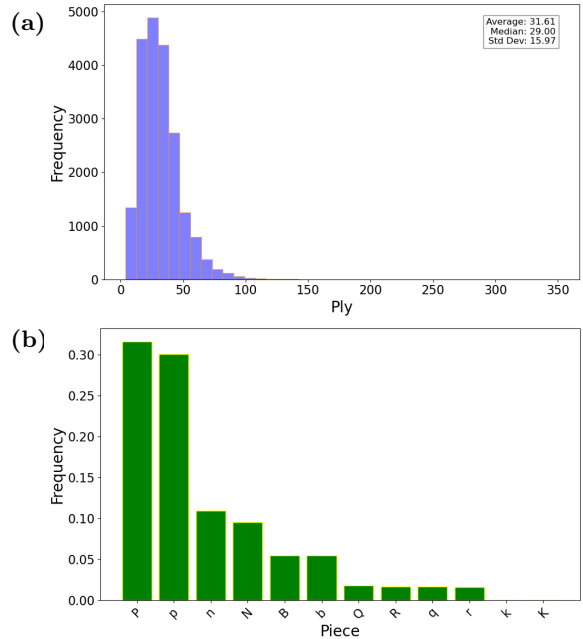
\begin{figure}[h!]
	\centering
	\begin{picture}(0,0)
		\put(-10,110){\textbf{(a)}} 
	\end{picture}
	\includegraphics[width=0.4\textwidth]{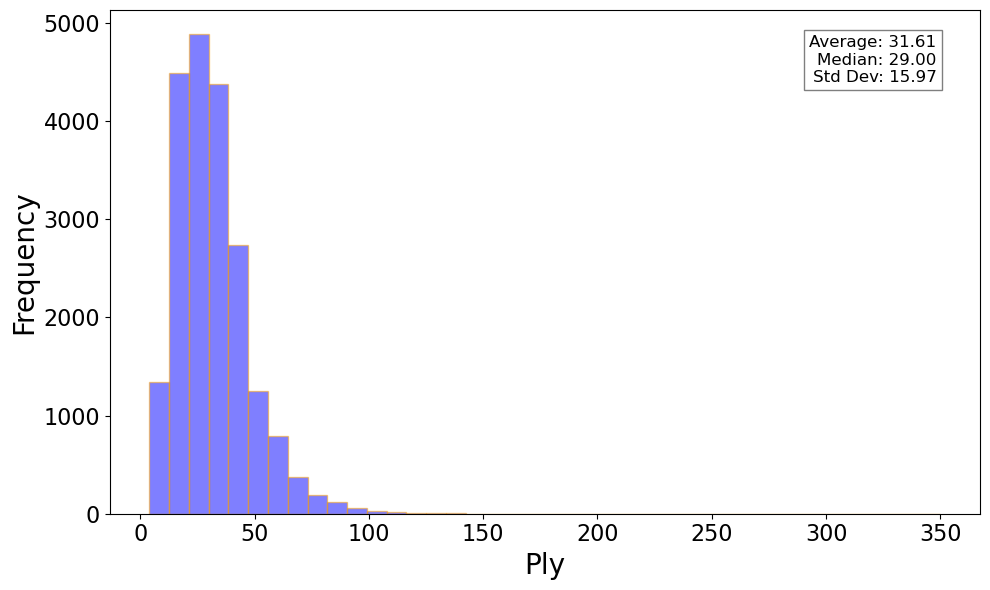}
	\begin{picture}(0,0)
		\put(-220,-10){\textbf{(b)}} 
	\end{picture}
	\includegraphics[width=0.4\textwidth]{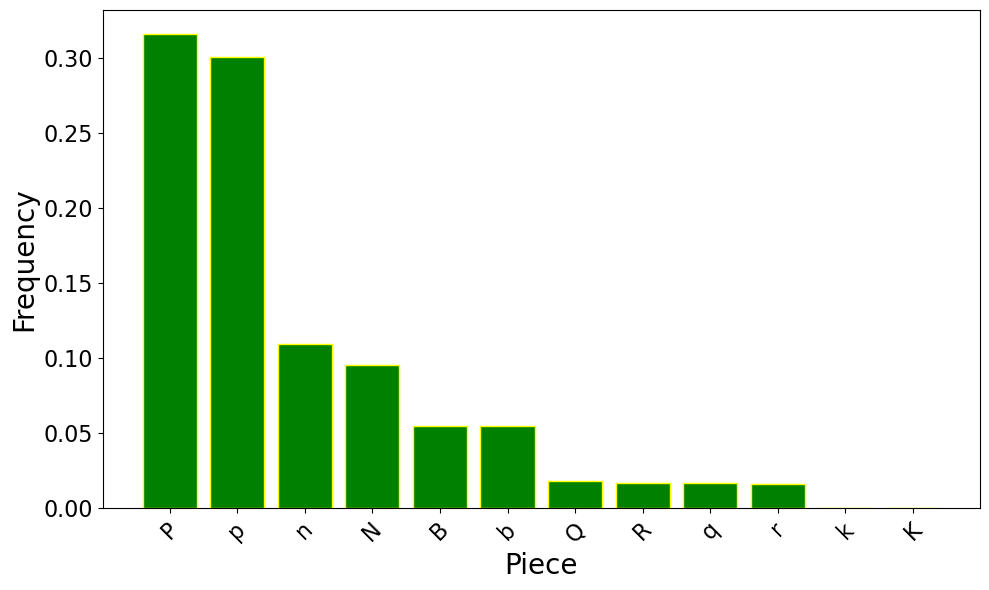}
	\caption{{\bf Properties of the maximum fragility score}. (a) Histogram of the ply at which the fragility score is maximum. On average, the maximum is at ply $\approx 32$ (which corresponds to the move $16$). (b) Histogram of the key piece (attacked piece with the largest BC). We observe that in about $60\%$ of the games the pawns were the key pieces, and for $20\%$ the knights are the key pieces. These results are obtained for a total of $20,685$ games for a selection of world top players.}
	\label{fig:3}
\end{figure}

For each game, we compute the fragility score, identify its maximum and the corresponding ply and key piece (attacked piece with the largest betweenness centrality). We first show the histogram of the ply at which the maximum occurs (Fig.~\ref{fig:3}(a)). Typically, we see that
the most fragile positions arise around ply $32$ (which correspond to the $16^{th}$ move, or more generally to about $30\%$ of the total number of plies). The histogram for the key pieces (Fig.~\ref{fig:3}(b)) shows that the key pieces involved in fragile positions are mostly pawns ($\approx 60\%$ of cases), followed by  knights ($\approx 20\%$).

\subsection{Fragility and Stockfish evaluation}

\begin{figure}[h!]
	\centering
	\includegraphics[width=0.5\textwidth]{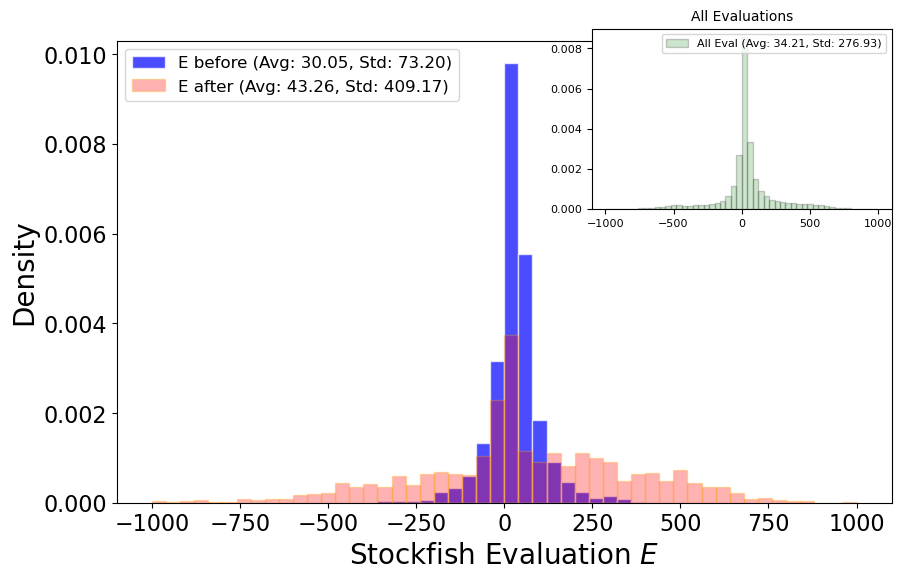}
	\caption{{\bf Evaluation before and after the maximum fragility.} We show the distribution of the Stockfish evaluation $E$ (in centipawn units) before (in blue) and after (in red) the maximuum fragility ply (and in the inset the full histogram for comparison). Before this ply, the evaluation is small with an average of $30$ (and a small dispersion of order $\sim 70$), and large values of $E$ (either positive or negative) indicating a win or a loss for white are typically observed after this maximum fragility ply. These results were obtained by aggregating data from $100$ games for each player considered here, amounting to a total of $1,000$ games }
	\label{fig:4}
\end{figure}

It is interesting to explore whether there is a relationship between fragility and the game's evaluation by an engine such as Stockfish. Stockfish provides an evaluation $E$ (in centipawn units) that indicates which side is winning: a large positive $E$ suggests that White is winning, while a negative value indicates an advantage for Black. The value of $E$ can be interpreted as the equivalent gain or loss in terms of pawns. We present in Fig.~\ref{fig:4} the distribution of the Stockfish evaluation $E$ before and after the maximum fragility ply, with the full histogram shown in the inset for comparison. Before this ply, the evaluations are generally low, averaging around $30$ with a small dispersion of approximately $70$. Larger values of $E$, indicating a clear win or loss, are typically observed afterward. These results suggest that the maximum fragility ply can indeed be considered as a turning point where the moves done can decide of the fate of the game. 

\subsection{Universality of fragility}
\begin{figure}[h!]
	\centering
	\includegraphics[width=0.5\textwidth]{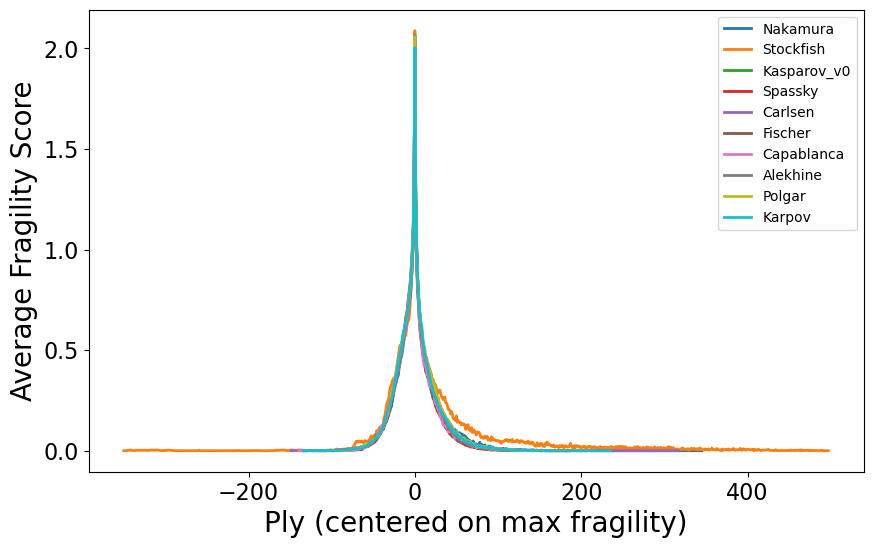}
	\caption{{\bf Average universal fragility score.} We show here the average aligned fragility score (we center the fragility score for each game around its maximum value and then perform the average over all games of the same player). Apart from small deviation observed for the engine Stockfish, all average fragility scores align almost perfectly.}
	\label{fig:5}
\end{figure}

For each game, we compute the total fragility score versus the ply, and align the fragility scores around the ply with the maximum fragility, centering the maximum for comparison across games. The average fragility score is then computed across all games, providing an aggregate measure of how fragility evolves. We average over all games of the same player (for a total of $20,685$ games and which mix a large variety
of different openings). The result is shown in Fig.~\ref{fig:5}.
We observe a surprising universality: the average fragility score is the same for all players and for all openings. Interestingly, we observe a slight 
difference for the Stockfish engine against other engines, which is probably related to the fact that games played by computers have a larger total number of moves.

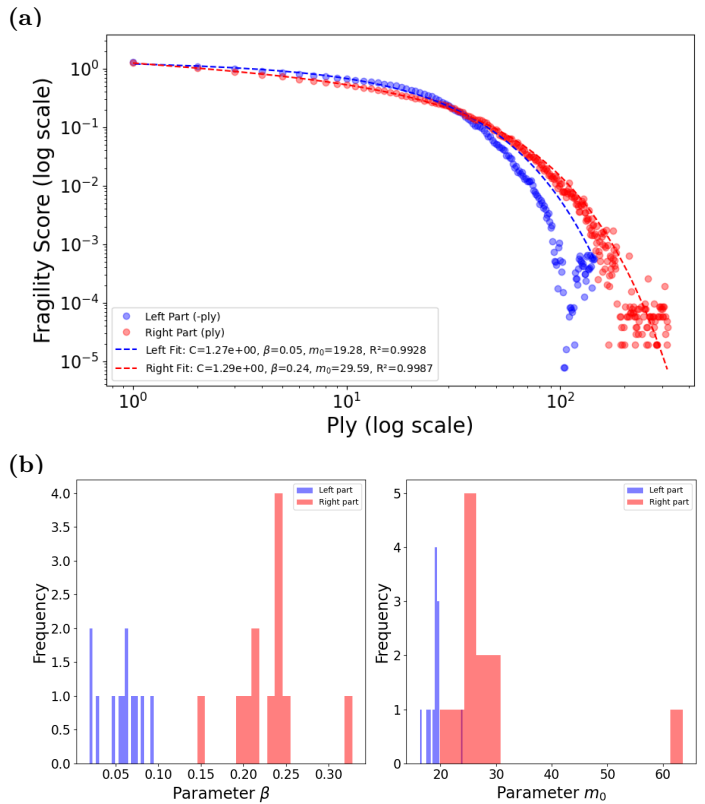
\begin{figure}[h!]
	\centering
	\begin{picture}(0,0)
		\put(-130,0){\textbf{(a)}} 
	\end{picture}
	\includegraphics[width=0.5\textwidth]{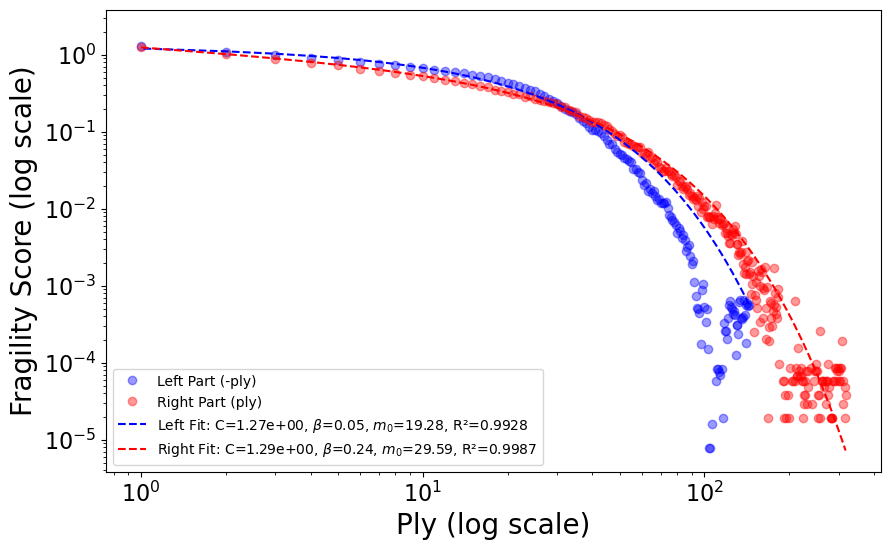}
	\begin{picture}(0,0)
		\put(-130,0){\textbf{(b)}} 
	\end{picture}
	\includegraphics[width=0.5\textwidth]{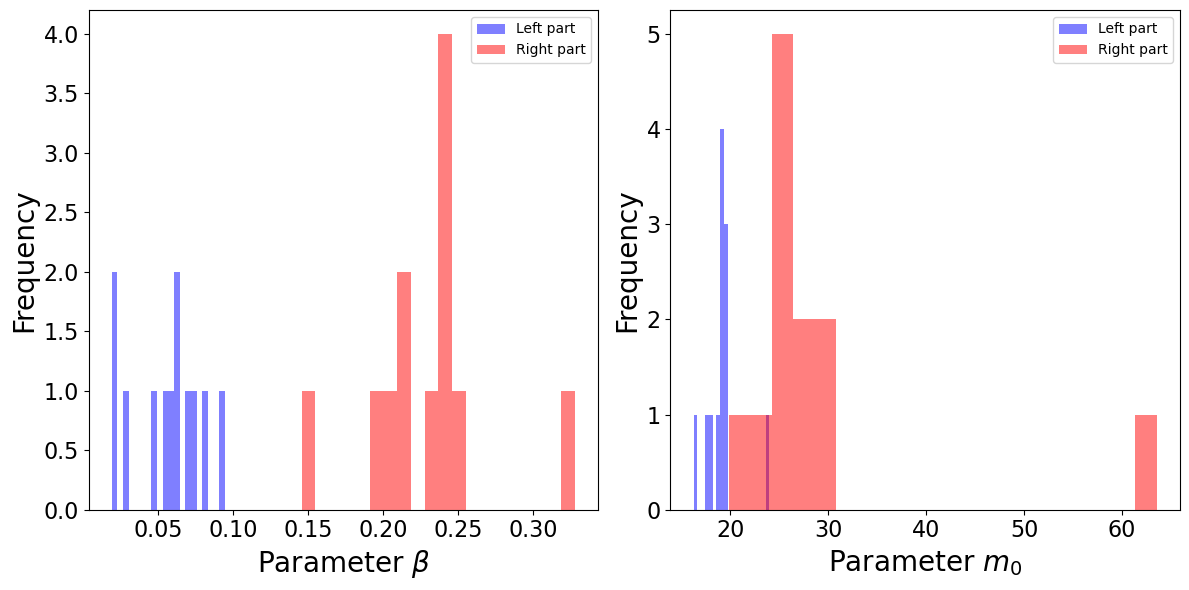}
	\caption{{\bf Fitting the average universal fragility score.} (a) We show here in the case of Carlsen's games the fits with the function of Eq.~\ref{eq:fitfunct}. The fitting parameters are in this case $\beta_l\approx 0.05$, $\beta_r\approx 0.24$, and $m_{ol}\approx 19$, $m_{0r}\approx 30$ (in all cases $r^2>0.99$). (b) Histograms of the parameters $\beta$ and $m_0$ for the left and right parts for all games considered here. }
	\label{fig:6}
\end{figure}

In order to investigate this universal average fragility score, we split it in two parts. The left part captures the behavior of fragility
leading up to the maximum (i.e., before the maximum fragility ply), and
the right part that describes the decay of fragility after the maximum. Both the left and
right parts display a power law behavior with a sharp cut-off and we use the following fitting function
\begin{align}
  F(m)=\frac{C}{m^\beta} \mathrm{e}^{-m/m_0}
  \label{eq:fitfunct}
\end{align}
For example, we show this fit for Carlsen's games on the Fig.~\ref{fig:6}(a).
%
We use this function and fit all the average curves obtained for all  players considered here, and we find
on average for the left part $\beta_{l}=0.06\pm 0.022$ and $m_{0l}=19\pm 1.75$, and for the right part $\beta_r=0.23\pm 0.04$,
$m_{0r}=29\pm 10.7$. The histogram of these values are shown in Fig.~\ref{fig:6}(b). We note here that the values of $m_0$ are
peaked around their average value, with the exception of Stockfish that represents an outlier with $m_{0l}=24$, and $m_{0r}=63$ ($r^2>0.98$),
values that are larger than for human players ($\approx 30$ half-moves). The fragility thus lasts much longer than in human games, showing that the
tension can be sustained much longer for an engine while human players will release the tension faster.

The power law exponents are relatively small which implies that the behavior of fragility is governed by the exponential cut-off. Basically, we thus observe on average an exponential of the form $\mathrm{exp}(-m/19)$ for the left part and a longer larger tail after the max fragiliy with a decay over almost $30$ half-moves. The build up of the fragility then takes on average $8$ moves, while fragility decays much slower with an average of $15$ moves, indicating that the later phase of the game maintain a degree of fragility, possibly indicating prolonged positional tension or vulnerability. On average, before the maximum fragility,  the game is `flat' in terms of its fragility dynamics,  players are building their positions with minimal risk or sharp changes. For the right part, the larger value \(m_0 = 29\) indicates that the decay of fragility occurs more slowly in the late game compared to the early game. This suggests that fragility persists for longer in the later phases, which is consistent with the idea that every move becomes more critical. In the endgame, small inaccuracies can have a more pronounced effect on the outcome, but fragility remains present over more moves before it decays completely. These results quantitatively describe the evolution of fragility during a chess game. The early game is stable with the relatively quick appearance of a large fragility, while the late game is more fragile and tactically rich, with each move having a more significant impact. This reflects typical chess dynamics where the opening is balanced and the endgame is much more critical in determining the outcome.

\begin{figure}[h!]
	\centering
	\begin{picture}(0,0)
		\put(-10,140){\textbf{(a)}} 
	\end{picture}
	\includegraphics[width=0.45\textwidth]{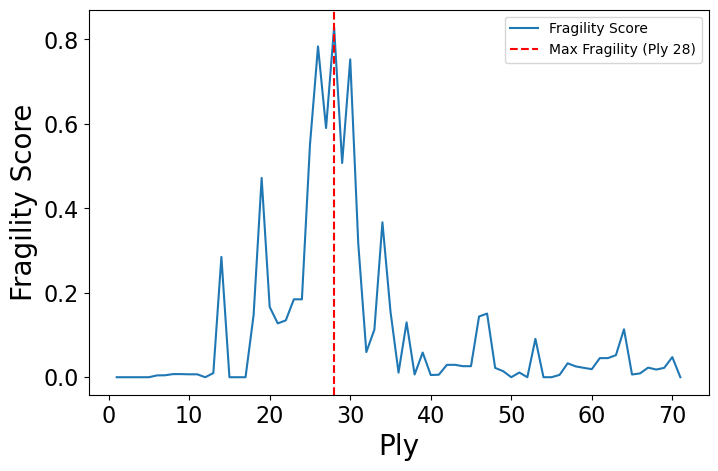}
	\begin{picture}(0,0)
		\put(-250,-10){\textbf{(b)}} 
	\end{picture}
	\includegraphics[width=0.45\textwidth]{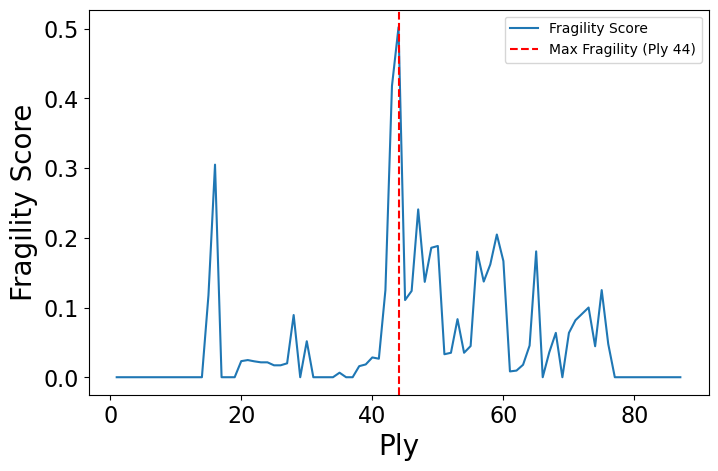} 
	\caption{{\bf Fragility scores for specific games.} (a) Game Carlsen-Giri (Wijk aan Zee 2022, 1-0), and (b) Kasparov-Topalov (Wijk aan Zee 1999, 1-0).}
	\label{fig:7}
\end{figure}
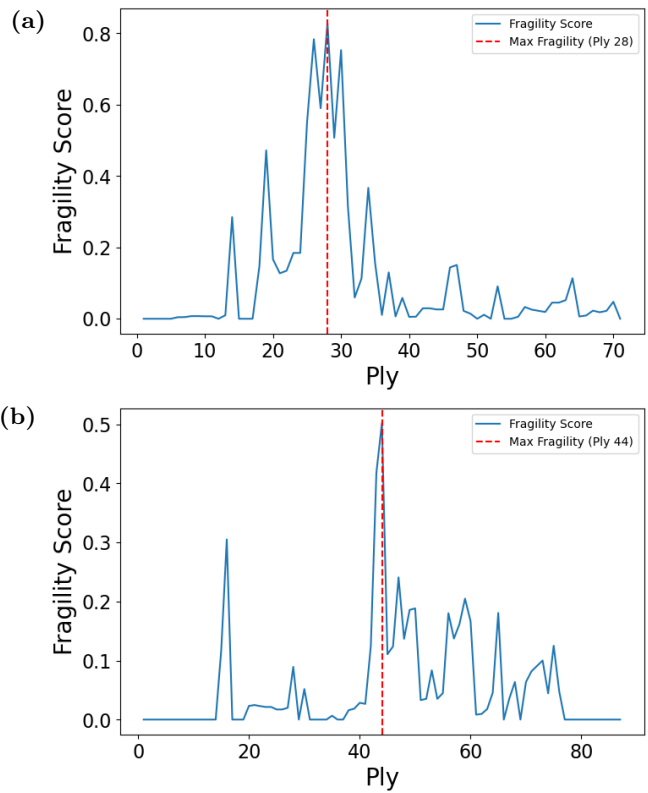

This universal curve is obtained by averaging over many games, but each game has specific fluctuations. We show in Fig.~\ref{fig:7}
two different games. For the first game (Carlsen-Giri 2022, 1-0), we indeed observe the average behavior with a gradual building of the fragility up to a maximum
followed by a gradual decrease. In contrast, for other games, such as Kasparov-Topalov (1999), the behavior is further away from this simple image and reflects the turbulent nature of this specific game with ferocious fight from both players with numerous tactical themes and a king hunt that led the king all the way to the other side of the board.
%
While these fluctuations may slightly reduce the practical significance of the average universal fragility curve, they underscore the importance of the fragility score in capturing the dynamic nature of a chess game.

\section{Discussion}

Despite recent advancements in AI and computational power, the scientific analysis of chess remains an evolving field. Two major breakthroughs have significantly enhanced our understanding of the game. First, the rise of AI engines surpassing human capabilities has revisited traditional chess theory, challenging long-held strategic principles and opening up new possibilities \cite{Sadler:2019}. Second, the availability of large game databases now allows for extensive statistical analysis, revealing novel patterns and insights across many games.

In this paper, we introduce a new approach to chess analysis by quantitatively measuring the fragility of positions using graph-theoretic concepts. The fragility score, based on the interaction graph of the pieces, captures the tension within a position, serving as an indicator of critical transitions or turning points. Our analysis reveals universal patterns across games, regardless of the player or opening. Specifically, we find that fragility builds up approximately $8$ moves before the most fragile position and remains elevated for about $15$ moves afterward. These results suggest that positional fragility follows in general a common trajectory, with tension peaking in the middlegame and dissipating toward the endgame. Interestingly, fragility scores in games played by the engine Stockfish (against other engines such as AlphaZero) show slight deviations from the observed universal patterns, hinting at differences in optimal play by AI compared to humans. Moreover, in famous games, the maximum fragility often coincides with pivotal moments, characterized by brilliant moves that decisively shift the balance of the game. We note that we have focused on the maximum fragility in this study; however, exploring the overall `fragility landscape' could be a compelling direction for future research, as it may uncover non-trivial aspects of game dynamics.

The consistent behavior of fragility scores across diverse games reflects the underlying structure of interaction graphs, which capture the strategic significance of piece relationships. The gradual decay of fragility indicates a prolonged state of tension, where small inaccuracies can dramatically shift the balance of power. This analysis highlights the complex dynamics of chess, where the interaction between attack and defense shapes the game's overall structure. Our study offers a new tool for understanding positional dynamics and provides a foundation for future research in chess analysis. By bridging traditional chess theory with quantitative methods, we aim to further refine our understanding of the game and its critical moments.\\

\section*{Acknowledgements}
I thank Nadia Blom for interesting discussions and very insighful comments.

\section*{Code availability}

The Python codes used to analyze the games and compute the fragility score are available upon request.\\

\section{Methods}

\subsection{Data}

The standard format for games is the Portable Game Notation (PGN) format, which includes metadata (e.g., date, location, opponent) and the moves in algebraic notation (e.g., \texttt{a2a4}, \texttt{b7b8}).

We used publicly available data from online resources for chess games, in particular from \url{https://www.pgnmentor.com} that proposes
the games (in pgn format) for each player \cite{pgnmentor} or sorted by openings.

In particular, we discuss in this paper the game Mehedlishvili - Van Forrest (0-1) played at the 44th FIDE Chess Olympiad Chennai 2022 Open, and the games Carlsen-Giri (Wijk aan Zee 2022, 1-0), and the famous game Kasparov-Topalov (Wijk aan Zee 1999, 1-0). 
We also studied the top 10 games as defined by the chess.com website \cite{top10}. The pgn files for individual games (and also those for Stockfish against other engines, 2018-2021) can be found on the chessgames.com website \cite{chessgames}.

\subsection{Software}

We used the python-chess library (v1.9.4) for move generation, validation, and analysis \cite{pythonchess}. This library includes a Stockfish class for easy integration with the Stockfish chess engine \cite{pypisf}.


\end{document}